\documentclass[a4paper,11pt]{article}
\usepackage{ifpdf}
\ifpdf
  \usepackage[pdftex]{graphicx}
  \DeclareGraphicsExtensions{.mps,.pdf,.jpg,.png}%
\else
  \usepackage[dvips]{graphicx}
  \DeclareGraphicsExtensions{.mps,.eps}%
\fi

\usepackage{amsmath,amssymb,url}

\usepackage{bm}

\providecommand{\abs}[1]{\left\lvert#1\right\rvert}
\renewcommand{\phi}{\varphi}
\renewcommand{\Phi}{\varPhi}
\renewcommand{\Gamma}{\varGamma}
\renewcommand{\Lambda}{\varLambda}
\renewcommand{\Re}{\operatorname{Re}}

\renewcommand{\vec}[1]{\bm{#1}}
\renewcommand{\i}{\,\mathrm{i}}
\newcommand{\e}{\,\mathrm{e}}
\newcommand{\scm}{Schwarz--Chris\-tof\-fel mapping}
\newcommand{\R}{\mathbb R}

\newcommand{\pd}[2]{\dfrac{\partial#1}{\partial#2}}

\renewcommand{\Lambda}{\varLambda}
\renewcommand{\Delta}{\varDelta}

\renewcommand{\Phi}{\varPhi}

\newcommand{\Phiin}{\vec\Phi^{\text{in}}}

\newcommand{\Ttot}{T^+_{\text{tot}}}
\newcommand{\Rtot}{R^+_{\text{tot}}}

\begin{document}
\title{Fourier Methods for Harmonic Scalar Waves in General
  Waveguides}

\author{Anders Andersson\\School of Engineering, J\"onk\"oping
  University 
  \and B\"orje Nilsson\\Department of Mathematics, Linn{\ae}us 
  University
  \and Thomas Biro\\School of Engineering, J\"onk\"oping University}
\maketitle

The final publication is available at Springer via\\
 \url{http://dx.doi.org/10.1007/s10665-015-9808-8}

\begin{abstract}
  A set of semi-analytical techniques based on Fourier analysis is
  used to solve wave scattering problems in variously shaped
  waveguides with varying normal admittance boundary conditions. Key
  components are newly developed conformal mapping methods, wave
  splitting, Fourier series expansions in eigen-functions to
  non-normal operators, the building block method or the cascade
  technique, Dirichlet-to-Neumann operators, and reformulation in
  terms of stable differential equations for reflection and
  transmission matrices. For an example the results show good
  correspondence with a finite element method solution to the same
  problem in the low and medium frequency domain. The Fourier method
  complements finite element analysis as a waveguide simulation tool.
  For inverse engineering involving tuning of straight waveguide parts
  joining complicated waveguide elements, the Fourier method is an
  attractive alternative including time aspects. The prime motivation
  for the Fourier method is its added physical understanding primarily
  at low frequencies.
\end{abstract}

\section{Introduction}
\label{sec:intro}

Scattering of waves in guides of different shapes is a classical
problem in applied mathematics. The problem appears in many
applications: in acoustics and electrodynamics
\cite{jones1986,jin:2010}, quantum physics
\cite{londegan+carini+murdock:1999}, and water waves
\cite{abbott:1956,aimen+michelle+theodore:1998}. Water waves require
usually non-linear modelling whereas linear models are often adequate
for acoustics and electrodynamics.

When treating this problem mathematically, the task is to find
solutions to the wave equation in some of its forms, and an essential
part of this task for linear waves is to solve Helmholtz equation
\begin{equation}
  \label{eq:Helmholtz1}
  (\nabla^2+k^2)\phi=0
\end{equation}
in different geometries and with various boundary conditions.

For simple geometries, for example straight waveguides with hard or
soft walls, the solutions are easily found in terms of Fourier series
that is numerically efficient, except for high frequencies, or more
precisely, with the exception of situations where the width of the
waveguide is much larger than the wavelength. Examples using this
technique are found in basic text books on partial differential
equations. Straight waveguides with hard or soft walls having abrupt
changes in geometry like bifurcations, $90^{\tiny{\rm o}}$ bends and
open ends can be dealt with using Fourier methods such as Wiener-Hopf
and mode matching techniques \cite{mittralee1971}. By combining these
techniques with the Building Block, BB, method
\cite{nilssonbrander1981b}, also denoted as the generalized scattering
matrix \cite{mittralee1971} and the cascade \cite{jones1986}
technique, it is possible to consider sudden area changes
\cite{mittralee1971,nilssonbrander1981b}. Except for simple waveguide
elements like circular bends, see e.g. \cite{bironilsson2005}, the
treated elements are non-smooth but even for the circular bend, its
connection to adjoining waveguides are modelled as non-smooth
\cite{bironilsson2005}. Generalizations to more complicated boundary
conditions, like normal surface impedance or admittance, have also
been done \cite{buyukaksoycinar2005}.

The Finite Element method (FE), was originally rarely used for
waveguide problems. However, due to evolution of both methods and
computers and of a big supply of commercial software, FE is today often
the natural choice, even for these problems
\cite{zienkiewicz+taylor+zhu:2008,Ihlenburg:1998}.

Nevertheless, there is a long tradition in using analytic or
semi-analytic methods for the solution of two-dimensional static
problems, like the harmonic or the bi-harmonic equations, with a
complicated geometry using conformal mappings
\cite{muskhelishvili:1962}. Such conformal mappings have also been
used to improve finite element and finite difference methods to
achieve a more efficient and controlled mesh generation
\cite{ives+liutermoza:1977}. During recent years, numerically
efficient conformal mapping algorithms have been developed, first for
piecewise linear \cite{sctoolbox} and then for smooth boundaries
\cite{andersson-outpol:2008,andersson-acf:2009}. Examples of the
application of these or similar numerical conformal mapping methods
for waveguides are \cite{Andersson:2006,Andersson-Nilsson:2009} for
time harmonic acoustic and electromagnetic waves and
\cite{nachbin+daSilvaSimoes:2012} for non-linear time dependent water
waves. Here, \cite{Andersson:2006,Andersson-Nilsson:2009} are based on
numerically stable methods for propagation in straight waveguides with
varying parameters, see e.g. \cite{Fishman:1998,Nilsson:2002}, whereas
\cite{nachbin+daSilvaSimoes:2012} is based on a stabilizing iterative
method.

As mentioned above the BB method can be used for some generalizations
in geometry and to further improve the numerical performance, for the
semi-analytical methods in particular.

Parallel to this use of conformal mappings for internal propagation in
waveguides, quasi-conformal mappings have been used for acoustic and
electromagnetic cloaking in exterior domains
\cite{pendryschurigsmith2006}. Coordinate transformations that are not
orthogonal but simple to generate have also been used for
waveguides. Their common drawback with slow convergence for the
solution, due to that homogeneous boundary conditions are transformed
into inhomogeneous ones, has recently been removed by adding
supplementary modes \cite{maurel+mercier+felix:2014}.

Simulation tools for wave propagation have a wide range of
applications with varying requirements from the end user. Examples
from industrial requirements are:
\begin{enumerate}
\item For the analysis of a given design at the end of the design
  process, it is required that many effects, like complicated geometry
  and physics, are taken into account with a sufficient accuracy. At
  this stage, the calculation time has not the highest priority given
  that it is possible to perform the calculations.
\item In contrast, the accuracy is less important at an initial
  schematic parameter survey. However, the calculation time might be
  more critical and the influence of the parameters should be modelled
  with sufficient care.
\item Quick simulation tools are required in inverse engineering where
  a design with a specified performance is looked for. The latter can
  later be checked with a more accurate but slower method. Quick
  simulation tools are also required in real time applications even at
  the expense of reduced accuracy.
\item In the search for an optimal design in discussions with
  engineering colleagues it may be important to include physical
  interpretations of the simulation. The physical interpretation of
  the simulations is also important if it is required to generalize
  the mathematical models by using experimental results.

\end{enumerate}

A successful application of the combination of Fourier methods and the
BB method meeting requirement no 3 is the design of micro wave filters
\cite{bironilsson2005}. At that time, commercially available FE
softwares were not fast enough to be employed alone but were used to
check the found design. Even if FE solvers are much quicker today, the
BB method is still very attractive for inverse engineering involving
the variation of lengths of straight waveguide parts connecting fixed
more complicated waveguide parts. In addition, the BB method improves
the physical understanding, requirement 4 in particular if physical
modes are used making a natural connection to semi-analytical
methods. The need for having a good physical understanding is stressed
by \cite{nachbin+daSilvaSimoes:2012} and also that conformal mappings
add physical understanding to the solution with the numerical method.

The wide range of requirements from industry and further requirements
from more fundamental research indicate that more than one simulation
tool might be motivated. To this end it is interesting to investigate
if semi-analytical methods based on Fourier analysis can complement FE
and FD (finite difference) methods in the simulation of wave models
with complicated geometry and general normal impedance or admittance
boundary conditions. The prime motivation for these semi-analytical
methods, in the sequel called Fourier methods, is the increased
physical understanding. Even if it is exceptional, there are cases
according to the example above when the Fourier method is the
quickest one. Another case with an advantage for the Fourier
method is when the domain is naturally described with basis functions
having a non-discrete spectrum. One example is the radiation from
open pipes, which is conveniently solved with Wiener-Hopf methods, and
is naturally linked to other scattering wave guide elements with the
BB method. There are even cases when there is not yet a FE solution
available. One example is the convective instabilities that may exist
for waves in moving fluids \cite{nilssonbrander1980a} or plasmas
\cite{briggs1964}.

The purpose with the current paper is to demonstrate that Fourier
methods can solve time harmonic scattering problems in waveguides with
complicated geometry and general normal impedance or admittance
boundary conditions. This generalization of Fourier methods deals with
added complications compared to previous works
\cite{maurel+mercier+felix:2014,Nilsson:2002}, which assume vanishing
normal admittance. Only waveguides with smooth boundaries are
considered. Additional well-known tools such as mode matching and
Wiener--Hopf methods suitable for handling discontinuities, see for
example \cite{jones1986}, are necessary to solve general waveguide
problems, but are not used in this article. The high frequency regime,
for which asymptotic methods exist, is not included.

For the Fourier method, we use a toolbox containing a set of
methods:
\begin{itemize}
\item The BB method makes it possible to divide a complicated geometry
  into several tractable parts.
\item Different conformal mapping methods are used to further simplify
  the geometry.
\item Reformulation of (\ref{eq:Helmholtz1}) assuming that the field
  can be expressed in Fourier series.
\item Determination of numerically stable differential equations for
  reflection and transmission matrices means that these can be
  determined for each part of the waveguide.
\item Dirichlet-to-Neumann operators make it possible to formulate and
  solve numerically stable differential equations for the field in the
  waveguide.
\end{itemize}

In an example where the field in the waveguide is scalar, we show how
these techniques can be combined, in order to get the complete
solution of Helmholtz equation in a two-dimensional waveguide with
general geometry and normal admittance boundary conditions. In order
to check the accuracy of the method, results are presented not only in
the low frequency regime.  The results in this example are then
compared with the results when using commercial FE software to solve
the problem.

The plan of the paper is as follows: In Section~\ref{sec:prel}, the
mathematical basis for a Fourier solution is outlined and in
Section~\ref{sec:oneblock}, it is described how the field as well as
reflection and transmission matrices are determined in a single
``block'', using conformal mappings and different reformulations of
equation~(\ref{eq:Helmholtz1}).  Section~\ref{sec:comb-blocks-build}
shows how these blocks can be combined, using the BB
method. Section~\ref{sec:numerical-example} contains the example
problem and a detailed description of the techniques used to solve
it. Finally, a discussion and some conclusions are included in
Section~\ref{sec:conclusion}.

\section{Formulation of the scattering problem}
\label{sec:prel}

The problem under investigation is scattering of scalar harmonic waves
$\phi$ in a two-dimensional waveguide $V$ of rather general
shape. $\phi$ solves Helmholtz equation%
\begin{equation}
  (\nabla^{2}+k^{2})\phi(\bm{r})=0,\quad\bm{r}\in V, \label{101}%
\end{equation}
in the interior $V$ of the waveguide and fulfils a homogeneous
boundary condition%
\begin{equation}
  \dfrac{\partial \phi}{\partial n}=\text{i}k\beta \phi,
  \text{ }\bm{r}\in\partial V\text{,} \label{102}%
\end{equation}
on the boundary $\partial V$. Here, $k\in\R^{+}$ is the wave
number, $\beta\in\mathbb{C}$, $\Re\beta\geq0$ is the (normalised)
normal surface admittance and $\widehat{\bm{n}}$ is the outward
pointing normal to $\partial V$. Special cases of the boundary
conditions are $\beta=0$ (Neumann or hard) and $\beta=\i\infty$
(Dirichlet or soft). The formulation assumes the time dependence
$\exp(-\i kc_{0}t)$ in the underlying wave equation where $c_{0},$ is
the wave speed of the medium in the waveguide.  In the formulation
(\ref{101}-\ref{102}), physical dimensions are omitted.  Sources in
terms of incident waves will be specified at the end of this section.

\begin{figure}[t]
  \centering
  \includegraphics[scale=0.8]{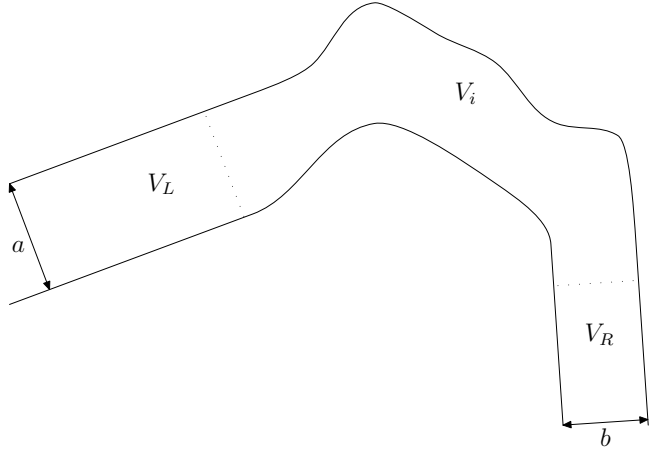}
  \caption{Waveguide $V=V_L\cup V_i\cup V_R$ consisting of straight
    parts $V_L$ and $V_R$ open to infinity and a bounded connecting
    part $V_i$.}
  \label{fig:wg1}
\end{figure}

The waveguide $V$ consists of three parts: an inner bounded part
$V_{\text{i}}$ and two straight semi-infinite parts $V_{\text{L}}$ and
$V_{\text{R}}$. An axial co-ordinate $u$ is associated to the
waveguide with the positive axis aligned with $V_{\text{R}}$ and the
negative with $V_{\text{L}}.$ The transverse coordinate in the
straight waveguides $V_{\text{L}}$ and $V_{\text{R}}$, orthogonal to
$u$, is called $v$.  Fig.  \ref{fig:wg1} depicts this waveguide
schematically.%
Using the notation $\mathbb{R}_{u_0}^{\pm}=\left\{ u\in\mathbb{R}:\
  u\gtrless u_{0}\right\} ,$ the straight waveguide parts are
$V_{\text{L}%
}=[0,a]\times\mathbb{R}_{u_{\text{L}}}^{-}$ and $V_{\text{R}%
}=[0,b]\times\mathbb{R}_{u_{\text{R}}}^{+}$.  For simplicity it is
assumed that the surface admittance $\beta=0$ on $\partial
V_{\text{L}}\cap\partial V$ and on $\partial V_{\text{R}}\cap\partial
V$, whereas $\beta$ can be non-vanishing and varying on
$V_{\text{i}}.$

As a preparation for the definition of the scattering problem, we
discuss propagation of waves in an infinite straight waveguide
$V_{\gamma}=[0,\gamma]\times\mathbb{R},$ representing $V_{\rm L}$ if
$\gamma=a$ or $V_{\rm R}$ if $\gamma=b$. If $\phi$ solves (\ref{101}) in
$V_{\gamma},$ $\phi$ can be uniquely splitted into
\begin{equation}
  \phi(u,v)=\phi_{-}(u,v)+\phi_{+}(u,v), \label{103}
\end{equation}
where%
\begin{equation}\label{103.2}
  \phi_{\pm}(u,v)=\sum_{n=0}^{\infty}\phi_{n}^{\pm}(u)\psi
  _{\gamma n}(v),0\leq v\leq \gamma.
\end{equation}
Here, $\psi _{\gamma n}(v)$ is a basis function, $\phi_{n}^{\pm}(u)$
is the corresponding expansion coefficient,
\begin{equation}
  \left\{
    \begin{array}
      [c]{l}%
      \phi_{n}^{\pm}(u)=A_{n}^{\pm}\text{e}^{\pm\text{i}\alpha_{\gamma n}u}\\[1ex]%
      \psi
      _{\gamma n}(v)=\cos\frac{n\pi v}{\gamma}
    \end{array}
  \right.  , \label{103.4}%
\end{equation}
$A_{n}^{\pm}$ is a constant modal amplitude and
\begin{equation}
  \alpha_{\gamma n}=\left\{
    \begin{array}
      [c]{l}%
      \sqrt{k^{2}-(n\pi/\gamma)^{2}},\text{ }k\geq n\pi/\gamma\\[1ex]%
      \text{i}\sqrt{(n\pi/\gamma)^{2}-k^{2}},\text{ }k<n\pi/\gamma%
    \end{array}
  \right.   \label{106}%
\end{equation}
is the axial wavenumber.

If $\phi_{+}\,\not\equiv\, 0,$ we say that there is a source at
$u=-\infty,$ and if $\phi_{-}\,\not\equiv\, 0,$ there is a source at
$u=+\infty.$ It makes it natural to denote $\phi_{-}$ or
$\phi_{n}^{-}$ as left-going and $\phi_{+}$ or $\phi_{n}^{+}$ as
right-going.

In straight waveguide parts like $V_{\text{L}}$ and $V_{\text{R}}$ it
is possible to identify the right-going wave $ \phi_{+}(u,v)$ with its
coefficient vector $
\bm{\phi}_{+}(u)=(\phi_{0}^{+}(u),\phi_{1}^{+}(u),\ldots)^{\text{T}}$
and $ \phi_{-}(u,v)$ with $
\bm{\phi}_{-}(u)=(\phi_{0}^{-}(u),\phi_{1}^{-}(u),\ldots)^{\text{T}}$. This
coefficient formulation is now used to formulate the scattering
problem where the so-called incident waves, the right-going or plus
wave in $V_{\text{L}}$ and the left-going or minus wave in
$V_{\text{R}}$, are known representing sources in the far left and
right portions of the waveguide, respectively. To be determined are
the remaining parts of the waves in $V_{\text{L}}$ and $V_{\text{R}}$,
denoted the scattered waves. Due to linearity it is possible to relate
linearly the known incident waves $\bm{\phi}_{\text{inc}}^{+}(u_1)$
and $\bm{\phi}_{\text{inc}}^{-}(u_2)$ with the scattered waves
$\bm{\phi}^{+}(u_2)$ and $\bm{\phi}^{-}(u_1)$ using
\begin{equation}
  \label{107}
  \begin{pmatrix}
    \vec \phi^+(u_2)\\
    \vec \phi^-(u_1)
  \end{pmatrix}= \mathcal{S}(u_1,u_2)
  \begin{pmatrix}
    \vec \phi_{\text{inc}}^+(u_1)\\
    \vec \phi_{\text{inc}}^-(u_2)
  \end{pmatrix},\, u_1\, {\rm in}\, V_{\rm L}\, {\rm and}\, u_2\, {\rm
    in}\, V_{\rm R}.
\end{equation}
Since the scattered waves are formed from a transmitted and a
reflected part, it is convenient to write the scattering matrix $
\mathcal{S}(u_1,u_2)$ as
\begin{equation}
  \label{108}
  \mathcal{S}(u_1,u_2)=
  \begin{pmatrix}
    T^+(u_2,u_1)& R^-(u_1,u_2)\\
    R^+(u_2,u_1)& T^-(u_1,u_2)
  \end{pmatrix}
  .
\end{equation}
Here, $R^-(u_1,u_2)$ and $R^+(u_2,u_1)$ are reflection matrices, and
$T^+(u_2,u_1)$ and $T^-(u_1,u_2)$ are transmission matrices.

The scattering matrix for an infinite straight waveguide of width $a$
is
\begin{equation}
  \label{109}
  \mathcal{S}(u_1,u_2)=
  \begin{pmatrix}
    U_a(u_2-u_1)&0\\
    0&U_a(u_1-u_2)
  \end{pmatrix}
  .
\end{equation}
where
\begin{equation}
  \label{eq:S}
  U_a(u)=
  \begin{pmatrix}
    \e^{\i\alpha_{a0}u}&0&0&\cdots\\
    0&\e^{\i\alpha_{a1}u}&0&\cdots\\
    0&0&\e^{\i\alpha_{a2}u}&\\
    \vdots&\vdots&&\ddots
  \end{pmatrix}.
\end{equation}

All matrices are infinite in size. In numerical calculations, the
matrices are truncated. Note, however, that it may be required to
include cut-off modes in the calculations.

If the wave field is required in only the straight waveguides
$V_{\text{L}}$ and $V_{\text{R}}$, the solution to the above
scattering problem is equivalent to the scattering matrix
$\mathcal{S}(u_1,u_2)$. This scattering matrix approach facilitates
understanding and formulation as well as reduces the computational
time of scattering problems. The reason is that the scattering matrix
for a complicated structure can be synthesized from the scattering
matrices for simple structures as will be shown in section 4.

\section{Solution of the one-block problems}
\label{sec:oneblock}

Wave scattering in a complicated geometry with varying boundary
conditions can be treated as a series of simpler problem, using the so
called Building Block Method, see
Section~\ref{sec:comb-blocks-build}. The method determines reflection
and transmission matrices for the waveguide, given that these matrices
have been determined for each part (``block'') of the waveguide.

In this section, we show how reflection and transmission matrices for
a single block are established, but also, using a Dirichlet-to-Neumann
formulation, how the wave field inside the block could be determined
if required. It is assumed that the scattering matrix for each block
is determined for anechoic terminations. Otherwise, the interactions
due to the termination should have been included in the Building Block
Method making it much more complicated. To this end each block is
assumed to be an infinitely long waveguide with parallel straight
walls and constant boundary conditions outside some bounded transition
region. Virtual intermediate straight waveguides, required by the
anechoic condition, but with a vanishing length can be inserted
\cite{nilssonbrander1981b} demonstrating that the anechoic termination
assumption introduces in practice no limitation of the method.

In each such geometry, the boundary value problem
\begin{equation}
  \label{eq:bvp1}
  \begin{cases}
    \left(\nabla^2+k^2\right)\phi(x,y)=0&\text{in the waveguide,}\\[1ex]
    \pd \phi n=\i k\beta(t)\phi&\text{on the boundary,}
  \end{cases}
\end{equation}
where $\beta$ varies smoothly with some boundary parameter $t$, should
be solved. For the sake of simplicity, we assume that one of the
waveguide walls is hard, giving a Neumann boundary condition there.

\begin{figure}[t]
  \centering
  \includegraphics[scale=1]{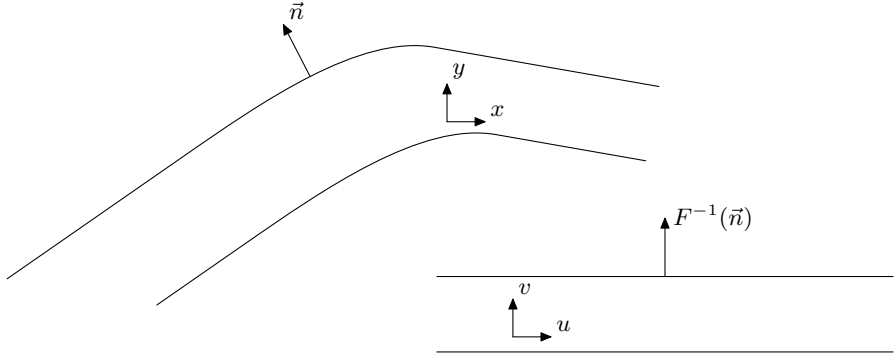}
  \caption{A single block in the $z=x+\i y$ plane and the $w=u+\i v$
    plane.}
  \label{fig:confmap}
\end{figure}

We use a conformal mapping
\begin{equation*}
  F:w=u+\i v\to z=x+\i y
\end{equation*}
to transform the geometry into a straight horizontal waveguide
\mbox{$\{u\in\R,0\le v\le1\}$} in the $(u,v)$-plane, see
Fig.~\ref{fig:confmap}. After this transformation, Eq.~(\ref{eq:bvp1})
yields
\begin{equation}
  \label{eq:bvp2}
  \begin{cases}
    \left(\nabla^2+k^2\mu(u,v)\right)\Phi(u,v)=0\\[1ex]
    \left.\pd{\Phi(u,v)}v\right|_{v=1}=\i kY(u)\Phi(u,1)\\[1.5ex]
    \left.\pd{\Phi(u,v)}v\right|_{v=0}=0
  \end{cases},
\end{equation}
where $\mu(u,v)=\abs{F'(w)}^2$ and $Y(u)=\beta(u)\abs{F'(u+i)}$. Note
that the boundary condition at the upper boundary, now at $v=1$, is
still a normal admittance boundary condition with an admittance that is
modified by the scale factor $\abs{F'(u+i)}$. Generally, the new
admittance is varying for constant $\beta$ but this variation can be
ignored asymptotically at both ends of the waveguide.

Following the techniques, outlined in \cite{Andersson-Nilsson:2009},
where an electromagnetic scattering problem is treated, or
\cite{Nilsson:2002}, where similar problems from acoustics are solved,
we expand $\Phi(u,v)$ in cosine Fourier series over $v$, assuming that
\begin{equation}
  \label{eq:fcosseries}
  \Phi(u,v)=\sum_n\Phi_n(u)\psi_n(u,v),
\end{equation}
where $\psi_n(u,v)=\cos(v\lambda_n(u))$ for functions
$\lambda_n(u),\quad n\in\mathbb N$.

The functions $\lambda_n$ are determined by the boundary condition
(\ref{eq:bvp2}), from which follows that $\lambda_n(u)$,
$n=0,1,\dots$, are solutions to the equation
\begin{equation}
  \label{eq:lambdaeq}
  \lambda_n(u)\tan\bigl(\lambda_n(u)\bigr)=-\i kY(u).
\end{equation}
By differentiating (\ref{eq:lambdaeq}), it follows that
\begin{equation}
  \label{eq:lambdaprim}
  \lambda_n'(u)=
  \frac{-\i kY'(u)}{Q(u)}
\end{equation}
and
\begin{equation}
  \label{eq:lambdabis}
  \lambda_n''(u)=-\i k\left(\frac{Y''(u)Q(u)-Y'(u)Q'(u)}
    {\bigl(Q(u)\bigr)^2}\right)
\end{equation}
where $(')$ stands for differentiating with respect to $u$ and
\begin{equation*}
  Q(u)=\tan\bigl(\lambda(u)\bigr)+
  \lambda(u)\Bigl(1+\tan^2\bigl(\lambda(u)\bigr)\Bigr).
\end{equation*}

From the expansions
\begin{align}
  \label{eq:vsinlambda}
  &v\sin(v\lambda_n(u))=\sum_m\alpha_{mn}(u)\cos(v\lambda_m(u)),\\
  \label{eq:vcos^2lambda}
  &v^2\cos(v\lambda_n(u))=\sum_m\beta_{mn}(u)\cos(v\lambda_m(u)),\\
  \label{eq:mucoslambda}
  &\mu(u,v)\cos(v\lambda_n(u))=\sum_m\mu_{mn}(u)\cos(v\lambda_m(u)),
\end{align}
follows the infinite-dimensional ordinary differential equation
\begin{equation}
  \label{eq:DE}
  \vec\Phi''(u)-A(u)\vec\Phi'(u)-B^2(u)\vec\Phi(u)=0,
\end{equation}
where the infinite vector $\vec\Phi=(\Phi_1 \Phi_2 \Phi_3\dots)^T$ and
the infinite matrices $A$ and $B^2$ have elements
\begin{equation}
  \label{eq:A}
  A_{mn}(u)=2\alpha_{mn}(u)\lambda_n'(u)
\end{equation}
and
\begin{equation}
  \label{eq:B2}
  B^2_{mn}(u)=\alpha_{mn}(u)\lambda_n''(u)
  +\beta_{mn}(u)\left(\lambda_n'(u)\right)^2
  +\delta_{mn}\left(\lambda_n(u)\right)^2
  -k^2\mu_{mn}(u).
\end{equation}

Note that $(\psi_m)$, with $\psi_m(v)=\cos v\lambda_m(u)$, is in
general not an orthogonal system for constant $u$ since the eigenvalue
problem (\ref{eq:bvp2}) is a regular Sturm-Liouville problem only if
$Y$ is purely imaginary. However, $(\psi_m)$ and
$(\overline{\psi_m})$, with a bar $\overline{\cdot}$ denoting the
complex conjugate, form a biorthogonal system meaning that the
bilinear form
\begin{equation}\label{eq:bilinear}
  \langle \psi_m,\psi_n\rangle
  =(\psi_m,\overline{\psi_n})
  =\int_0^1\psi_m(v)\psi_n(v){\rm d}v,
\end{equation}
vanishes if $n\neq m$. Here, for complex functions $f$ and $g$,
$(f,g)$ denotes the complex scalar product and the bilinear form
$\langle f,g \rangle$ is formally the same as the real scalar
product. This means that the expansion coefficients like
$\alpha_{mn}(u),\beta_{mn}(u)$ and $\mu_{mn}(u)$ in
(\ref{eq:vsinlambda}--\ref{eq:mucoslambda}), are found from
\begin{equation}\label{eq:am}
  a_m=\frac{\langle f,\psi_m \rangle}
  {\langle \psi_m,\psi_m \rangle}
\end{equation}
for the expansion
\begin{equation}\label{eq:expansion}
  f=\sum_m a_m \psi_m.
\end{equation}

Furthermore, the system $(\psi_m)$ is proved to be a basis in
L$^2(0,1)$ \cite{Schwartz:1954}. An alternative proof can be given
based on the existence of a sequence biorthogonal to $(\psi_m)$
using a theorem by Bari \cite{Bari:1944}, see \cite{Christensson:2003}
for a modern presentation of this theorem. It is interesting to note
that the biorthogonal property (\ref{eq:bilinear}) in a complex
Hilbert space with general complex $Y$ can be seen as an analytic
continuation of the orthogonality property in a real Hilbert space
with purely imaginary $Y$.

\subsection{Conformal mapping techniques}
\label{sec:confmap}

There are two indispensable requirements on the conformal mapping. The
first requirement is anechoic terminations at the ends of each
block. This is accomplished by treating the block as an infinite
waveguide which is straight and has constant cross-sections outside
some bounded region. We must therefore numerically construct a
conformal mapping from a straight infinite waveguide to an infinite
waveguide in which the walls at both ends are (at least)
asymptotically straight and parallel.

The second requirement relates to smoothness. To avoid singularities
in the matrices $A$ and $B^2$ in the differential equation
(\ref{eq:DE}), it follows from (\ref{eq:bvp2}) that the mapping must
have a bounded first derivative on the boundary, and additionally,
from (\ref{eq:lambdaprim}), (\ref{eq:lambdabis}), (\ref{eq:A}) and
(\ref{eq:B2}), bounded second and third derivatives if the boundary
has non-zero admittance.

In \cite{andersson-outpol:2008} and \cite{andersson-acf:2009},
conformal mapping techniques, suitable for this situation, are
developed. Both methods are built on the \scm, which guarantees that
the resulting waveguide walls are asymptotically straight and parallel
towards infinity, and they both result in regions with smooth boundary
curves, meaning that no singularities are introduced by the
mapping. In \cite{andersson-outpol:2008}, a suitable polygon,
surrounding the region under consideration, is determined, and the
conformal mapping is constructed using the \scm\ for that
polygon. In \cite{andersson-acf:2009}, the factors in a \scm\ are
replaced by so called approximate curve factors that round the corners
in a way that gives a smooth boundary curve.

\subsection{Re-transformation to stable equations}
\label{sec:stableeq}
The differential equation~(\ref{eq:DE}) cannot be solved directly by
numerical methods. However, there exist reformulations of
(\ref{eq:DE}) that are numerically stable.
In this section, we describe two such reformulations, built on two
different partitions of the wave field $\vec\Phi$.

Recall that the block is assumed to be an infinitely long waveguide
which is straight and has parallel hard boundaries outside some
central transition region. Let $\Omega_L$ and $\Omega_R$ be the
straight regions to the left and right respectively.  In $\Omega_L$
and $\Omega_R$, the matrix $A$ is zero, while $B$ is constant. Assume
that $B=B_-$ in $\Omega_L$ and $B=B_+$ in $\Omega_R$. In $\Omega_L$
and $\Omega_R$, $B^2$ is a real constant diagonal matrix, and to be
consistent with standard theory for straight waveguides, the square
roots of $B^2$ are chosen such that $B_-$ and $B_+$ have either
positive real or negative imaginary diagonal elements.

\subsubsection{Determination of the Reflection and Transmission
  matrices - The RT method}
\label{sec:RT}

Inspired by the partition $\phi=\phi_-+\phi_+$ in a straight waveguide
where the two terms can be seen as representing waves marching from
left to right and right to left respectively, we make the following
definition: Let for all $u\in\R$, the wave field
\begin{equation}
  \label{eq:Phipart1}
  \vec\Phi(u)=(\Phi_1(u)\ \Phi_2(u)\ \dots)^T
  =\vec\Phi^+(u)+\vec\Phi^-(u),
\end{equation}
where $\vec\Phi^+(u)$ and $\vec\Phi^-(u)$ represent waves marching to
the right and left respectively.

Let furthermore $C$ and $D$ be matrices, depending on $u$, such that
\begin{equation}
  \label{eq:plusminustodiff}
  \dfrac{\partial\vec\Phi}{\partial u}(u)=
  -C(u)\vec\Phi^+(u)+D(u)\vec\Phi^-(u),
\end{equation}
for all $u\in\R$. $C$ and $D$ can be defined in many different ways,
but they must be differentiable with respect to $u$, and since
(\ref{eq:DE}) must hold in $\Omega_L$ and $\Omega_R$ where $A(u)=0$,
it follows that $C=D=B_-$ in $\Omega_L$ and $C=D=B_+$ in
$\Omega_R$. We have used the definition
\begin{equation}
  \label{eq:CD}
  C(u)=D(u)=B_-+f(u)(B_+-B_-),
\end{equation}
where $f$ is a smooth function that is $0$ in $\Omega_L$ and $1$ in
$\Omega_R$.

Define, like in (\ref{108}), reflection and transmission matrices
$R^+$, $R^-$, $T^+$, $T^-$, such that for $u_1<u_2$,
\begin{equation}
  \label{eq:RT}
  \begin{pmatrix}
    \vec\Phi^+(u_2)\\
    \vec\Phi^-(u_1)
  \end{pmatrix}=
  \begin{pmatrix}
    T^+(u_2,u_1)&R^-(u_1,u_2)\\
    R^+(u_2,u_1)&T^-(u_1,u_2)
  \end{pmatrix}
  \begin{pmatrix}
    \vec\Phi^+(u_1)\\
    \vec\Phi^-(u_2)
  \end{pmatrix}.
\end{equation}
This means that $T^-$ and $R^-$ transmits respectively reflects the
left-going waves $\vec\Phi^-$, while $T^+$ and $R^+$ transmits
respectively reflects the right-going waves $\vec\Phi^+$.

From (\ref{eq:DE}), (\ref{eq:Phipart1}) and
(\ref{eq:plusminustodiff}), it is possible to derive, for details see
for example \cite{Nilsson:2002}, the equation
\begin{equation}
  \label{eq:diffphiplusminus}
  \pd{}u
  \begin{pmatrix}
    \vec\Phi^+\\\vec\Phi^-
  \end{pmatrix}
  =
  \begin{pmatrix}
    J&K\\
    L&M
  \end{pmatrix}
  \begin{pmatrix}
    \vec\Phi^+\\\vec\Phi^-
  \end{pmatrix},
\end{equation}
where
\begin{equation}
  \label{eq:alphabeta}
  \begin{split}
    &J=(C+D)^{-1}\left(-C'-B^2+(A-D)C\right),\\
    &K=(C+D)^{-1}\left(D'-B^2-(A-D)D\right),\\
    &L=(C+D)^{-1}\left(C'+B^2-(A+C)C\right),\\
    &M=(C+D)^{-1}\left(-D'+B^2+(A+C)D\right).
  \end{split}
\end{equation}

For the determination of $T^+$ and $R^+$, we consider (\ref{eq:RT})
assuming that there are no sources in $\Omega_R$. Let $u_2\in\Omega_R$
be constant and let $u=u_1$ vary. This means that $\vec\Phi^-(u_2)=0$,
and (\ref{eq:RT}) simplifies to
\begin{equation}
  \label{eq:RT2}
  \begin{cases}
    T^+(u_2,u)\vec\Phi^+(u)=\vec\Phi^+(u_2),\\
    R^+(u_2,u)\vec\Phi^+(u)=\vec\Phi^-(u).
  \end{cases}
\end{equation}
By differentiating (\ref{eq:RT2}) with respect to $u$, we get
\begin{equation}
  \label{eq:RTdiff1}
  \begin{cases}
    \pd{T^+}u(u_2,u)\vec\Phi^+(u)+T^+(u_2,u)\pd{\vec\Phi^+}u(u)=0,\\[1.5ex]
    \pd{R^+}u(u_2,u)\vec\Phi^+(u)+R^+(u_2,u)\pd{\vec\Phi^+}u(u)=
    \pd{\vec\Phi^-}u(u),
  \end{cases}
\end{equation}
and using (\ref{eq:diffphiplusminus}) and (\ref{eq:RT2}) once more,
the Ricatti equations
\begin{align}
  \label{eq:Ricatti1}
  \begin{split}
    \pd{R^+}u(u_2,u)&=-R^+(u_2,u)\bigl(J(u)+K(u)R^+(u_2,u)\bigr)\\
    &\qquad\qquad\qquad\qquad\qquad+L(u)+M(u)R^+(u_2,u)
  \end{split}\\
  \label{eq:Ricatti2}
  \pd{T^+}u(u_2,u)&=-T^+(u_2,u)\bigl(J(u)+K(u)R^+(u_2,u)\bigr)\\
  \intertext{follow. For $R^-$ and $T^-$, we proceed similarly
    assuming no sources in $\Omega_L$, and deduce the equations}
  \label{eq:Ricatti3}
  \begin{split}
    \pd{R^-}u(u,u_1)&=-R^-(u,u_1)\bigl(M(u)+L(u)R^-(u,u_1)\bigr)\\
    &\qquad\qquad\qquad\qquad\qquad+K(u)+J(u)R^-(u,u_1),
  \end{split}\\
  \label{eq:Ricatti4}
  \pd{T^-}u(u,u_1)&=-T^-(u,u_1)\bigl(M(u)+L(u)R^-(u,u_1)\bigr).
\end{align}
Using truncated matrices in place of $J$, $K$, $L$ and $M$, these
equations can be solved numerically with an ordinary differential
equation solver. (\ref{eq:Ricatti1}) and (\ref{eq:Ricatti2}) are
solved from right to left using $R^+(u_2,u_2)=0$ and $T^+(u_2,u_2)=I$
as initial values, while (\ref{eq:Ricatti3}) and (\ref{eq:Ricatti4})
are solved from left to right, using $R^-(u_1,u_1)=0$ and
$T^-(u_1,u_1)=I$ as initial values.

It is readily shown, see e.g. \cite{Nilsson:2002} for more details,
that the described procedure for calculating the transmission and
reflection matrices is numerically stable.

\subsubsection{Determination of the fields - The DtN method}
\label{sec:DtN}

To determine the field inside a single block, we reformulate
(\ref{eq:DE}) using Dirichlet-to-Neumann matrices, see also
\cite{Lu:1999} and \cite{Fishman:1998}. For this purpose, we make a
different partition of $\vec\Phi$. Let
\begin{equation}
  \label{eq:phiRphiL}
  \vec\Phi=\vec\Phi_R+\vec\Phi_L,
\end{equation}
where $\vec\Phi_R$ are waves with no sources to the right (in
$+\infty$) and $\vec\Phi_L$ are waves with no sources to the left (in
$-\infty$). Define Dirichlet to Neumann (DtN) matrices $\Lambda_R$ and
$\Lambda_L$ such that
\begin{align}
  \label{eq:DtNdef1}
  &\vec\Phi_R'(u)=-\Lambda_R(u)\vec\Phi_R(u),\\
  \label{eq:DtNdef2}
  &\vec\Phi_L'(u)=\Lambda_L(u)\vec\Phi_L(u).
\end{align}
$\vec\Phi_R$ and $\vec\Phi_L$ are both satisfying (\ref{eq:DE}), and
by differentiating (\ref{eq:DtNdef1}) and (\ref{eq:DtNdef2}), the
matrix equations
\begin{align}
  \label{eq:DtNDE1}
  &\Lambda_R'(u)=\bigl(A(u)+\Lambda_R(u)\bigr)\Lambda_R(u)-B^2(u)\\
  \label{eq:DtNDE2}
  &\Lambda_L'(u)=\bigl(A(u)-\Lambda_L(u)\bigr)\Lambda_L(u)+B^2(u)
\end{align}
follow. Since $A=0$ in $\Omega_L$ and $\Omega_R$,
\begin{align}
  \label{eq:LRDE1}
  &\vec\Phi_R'(u)+B_-\vec\Phi_R(u)=0,\quad
  \vec\Phi_L'(u)-B_-\vec\Phi_L(u)=0,&&u\in\Omega_L,\\
  \label{eq:LRDE2}
  &\vec\Phi_R'(u)-B_+\vec\Phi_R(u)=0,\quad
  \vec\Phi_L'(u)+B_-\vec\Phi_L(u)=0,&&u\in\Omega_R,
\end{align}
which means that if truncated matrices are used in place of $A$ and
$B^2$, (\ref{eq:DtNDE1}) and (\ref{eq:DtNDE2}) can be solved
numerically from right and left respectively, using the initial values
$\Lambda_R(u_2)=B_+$ and $\Lambda_L(u_1)=B_-$, where $u_1\in\Omega_L$
and $u_2\in\Omega_R$.  Finally, (\ref{eq:DtNdef1}) and
(\ref{eq:DtNdef2}) are solved numerically from left and right
respectively.

We have noted for our example that the Riccati
equations~(\ref{eq:DtNDE1}-\ref{eq:DtNDE2}) in the DtN method may be
stiff for certain values of $u$ but the stiffness is numerically
tractable. This kind of stiffness, not observed for
(\ref{eq:Ricatti1}-\ref{eq:Ricatti4}), is likely connected to a
singularity of $\Lambda_R(u)$ or $\Lambda_L(u)$ for a complex value
near the real $u-$axis, see for example \cite{Fishman:1998}. A
singularity on the real $u$ axis for $\Lambda_R$ can be avoided by
alternating between formulations for $\Lambda_R,\Lambda_R^{-1}$ and
$(\Lambda_R-dI)^{-1}$, where $d$ is real and $I$ is the identity
matrix, see \cite{lu+mclaughlin:2000}. Another alternative to handle
the singularities is to use recent numerical methods by which Riccati
matrix equations are integrated across singularities, even when no
knowledge about existence or location of these is at hand, see for
example \cite{Li-Kahan:2012}.

Numerical problems can also appear at a discrete set of frequencies
$k$ for methods based on field calculations, like the DtN method, when
trapped modes exist. In this case, the scalar field can for
mathematical reasons be determined only modulo an unknown factor times
the trapped mode. However, the scattering matrices can be determined
with the methods proposed in \cite{Nilsson:2002} also in the presence
of trapped modes.


\section{Combination of the Blocks - the Building
 Block Method}
\label{sec:comb-blocks-build}

The Building Block method (BB), see \cite{nilssonbrander1981b}, allows
the determination of reflection and transmission matrices for a
combination of several sections (``blocks'') of the waveguide, for
which these matrices are known.

\begin{figure}[t]
  \centering
  \includegraphics[width=\textwidth]{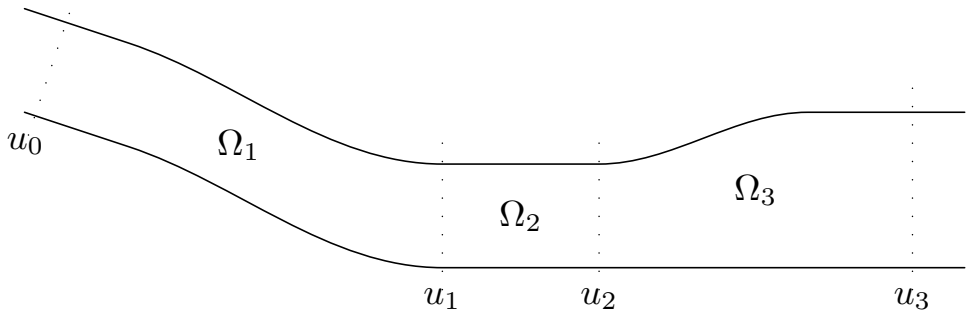}
  \caption{Waveguide divided in blocks. $\Omega_2$ is straight and
    with constant cross-section.}
  \label{fig:wg4}
\end{figure}

Assume that two subsequent blocks $\Omega_1$ and $\Omega_3$, are
connected by a region $\Omega_2$ which is straight with length $\ell$
and width $a$, see figure~\ref{fig:wg4}.  Furthermore, assume that
reflection and transmission matrices for $\Omega_1$ and $\Omega_3$ are
known.

To harmonize with the notation in the RT method, we choose here to
build on the expansion coefficients
$\vec\Phi(u)=\vec\Phi^+(u)+\vec\Phi^-(u)$ as varying with the position
along the waveguide rather than on modal amplitudes that are constant
in a straight waveguide. The latter procedure is the most common one
in the BB method. The reflection and transmission matrices are defined as
follows. Assume that a block begins and ends at position $u_j$ and
$u_{j+1}$ respectively and that there are no sources at the $u_{j+1}$
side of the block. If $T^+(=T^+(u_{j+1},u_j))$ is the transmission
matrix for waves $\vec\Phi^+(u)$ entering the block at $u_{j}$, then
$\vec\Phi^+(u_{j+1})=T^+\vec\Phi^+(u_{j})$ are the waves leaving the
block at position $u_{j+1}$. Let $u$ be a parameter, measuring the
distance along the central curve of the waveguide, and let $u=u_0$ at
the beginning (left end) of $\Omega_1$, $u=u_1=0$ at the border
between $\Omega_1$ and $\Omega_2$, $u=u_2=\ell$ at the border between
$\Omega_2$ and $\Omega_3$, assuming that the length of $\Omega_2$ is
$\ell$, and finally $u=u_3$ at the end of $\Omega_3$. Following the
definitions given in Eq.~(\ref{eq:RT}), we use the notation
\begin{align*}
  &R^+_1=R^+(u_1,u_0),&&T^+_1=T^+(u_1,u_0),\\
  &R^-_1=R^-(u_0,u_1),&&T^-_1=T^-(u_0,u_1),\\
  &R^+_3=R^+(u_3,u_2),&&T^+_3=T^+(u_3,u_2),\\
  &\Rtot=R^+(u_3,u_0),&&\Ttot=T^+(u_3,u_0).
\end{align*}

Assume that a right-marching field $\Phiin=\vec\Phi^+(u_0)$ is
entering $\Omega_1$ from the left, and that there are no sources to
the right of $\Omega_3$. Define matrices $C^\pm$ such that at the
border between $\Omega_1$ and $\Omega_2$, $\vec\Phi^+(0)=C^+\Phiin$
and $\vec\Phi^-(0)=C^-\Phiin$. Standard theory for straight waveguides
gives that in $\Omega_2$ at position $u$, the field is
\begin{equation}
  \label{eq:midfield}
  \vec\Phi(u)=(U_a(u)C^++U_a^{-1}(u)C^-)\Phiin,
\end{equation}
where $U_a$ is defined as in Eq.~(\ref{eq:S}) in
Section~\ref{sec:prel}.  Consequently, at the border between
$\Omega_2$ and $\Omega_3$, there are right-marching and left-marching
waves $U_a(\ell)C^+\Phiin$ and $U_a^{-1}(\ell)C^-\Phiin$ respectively.

Since there are no sources to the right of $\Omega_3$,
$U_a^{-1}(\ell)C^-=R_3^{+}U_a(\ell)C^+$, so
$C^-=U_a(\ell)R_3^{+}U_a(\ell)C^+$. But $C^+=T_1^{+}+R_1^{-}C^-$, and
hence
\begin{equation}
  \label{eq:ABRtotTtot}
  \begin{split}
    &C^+=\left(I-R_1^{-}U_a(\ell)R_3^{+}U_a(\ell)\right)^{-1}T_1^{+},\\
    &C^-=U_a(\ell)R_3^{+}U_a(\ell)C^+.
  \end{split}
\end{equation}

In summary the reflection and transmission matrices for the combined
waveguide $\Omega_1, \Omega_2$ and $\Omega_3$ is
\begin{equation}
  \label{eq:ABRtotTtot2}
  \begin{split}
    &\Ttot=T_3^{+}U_a(\ell)C^+,\\
    &\Rtot=R_1^{+}+T_1^{-}C^-.
  \end{split}
\end{equation}

The method has been known since the end of the 1940:s \cite{kerns1949}
and is denoted the Building Block Method \cite{nilssonbrander1981b} in
acoustic theory and cascade technique \cite{jones1986} in
electromagnetic theory.

\section{A numerical example}
\label{sec:numerical-example}
\begin{figure}[t]
  \centering
  \includegraphics[width=\textwidth]{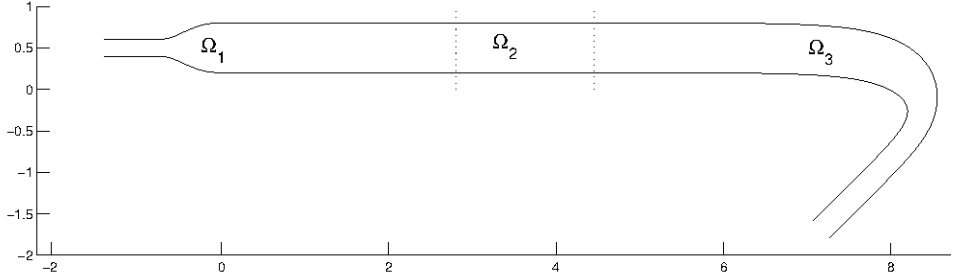}
  \caption{The waveguide in the example.}
  \label{fig:exwg}
\end{figure}
To illustrate the techniques, we solve the scattering problem
in the waveguide shown in Fig.~\ref{fig:exwg}.
The results, i.e., the total field in the waveguide given a known
field entering from the left, as well as reflection and transmission
operators for the waveguide, are calculated and compared with a finite
element method solution.

\subsection{Boundary conditions}
\label{sec:boundary-conditions}

The boundaries are hard, i.e., they have zero normal admittance except
for two intervals on the upper boundary. In both $\Omega_1$ and
$\Omega_3$, there is admittance at the upper boundary in the intervals
$F_j([-2,2]+\i)$ for $j\in\{1,2\}$. The admittance varies smoothly
along the border, and has a maximal level $\beta=0.5+0.5\i$ in the
intervals $F_j([-1,1]+\i)$, where the functions $F_j$ are the
conformal mappings defined in Section~\ref{sec:conformal-mappings}.

\subsection{Conformal mappings}
\label{sec:conformal-mappings}

The region is divided into three disjoint parts. $\Omega_1$ contains
the change in cross-section to the left, the middle section $\Omega_2$
is straight with constant cross-section, and $\Omega_3$ contains the
bending to the right.

\begin{figure}[t]
  \centering
  \includegraphics[width=0.45\textwidth]{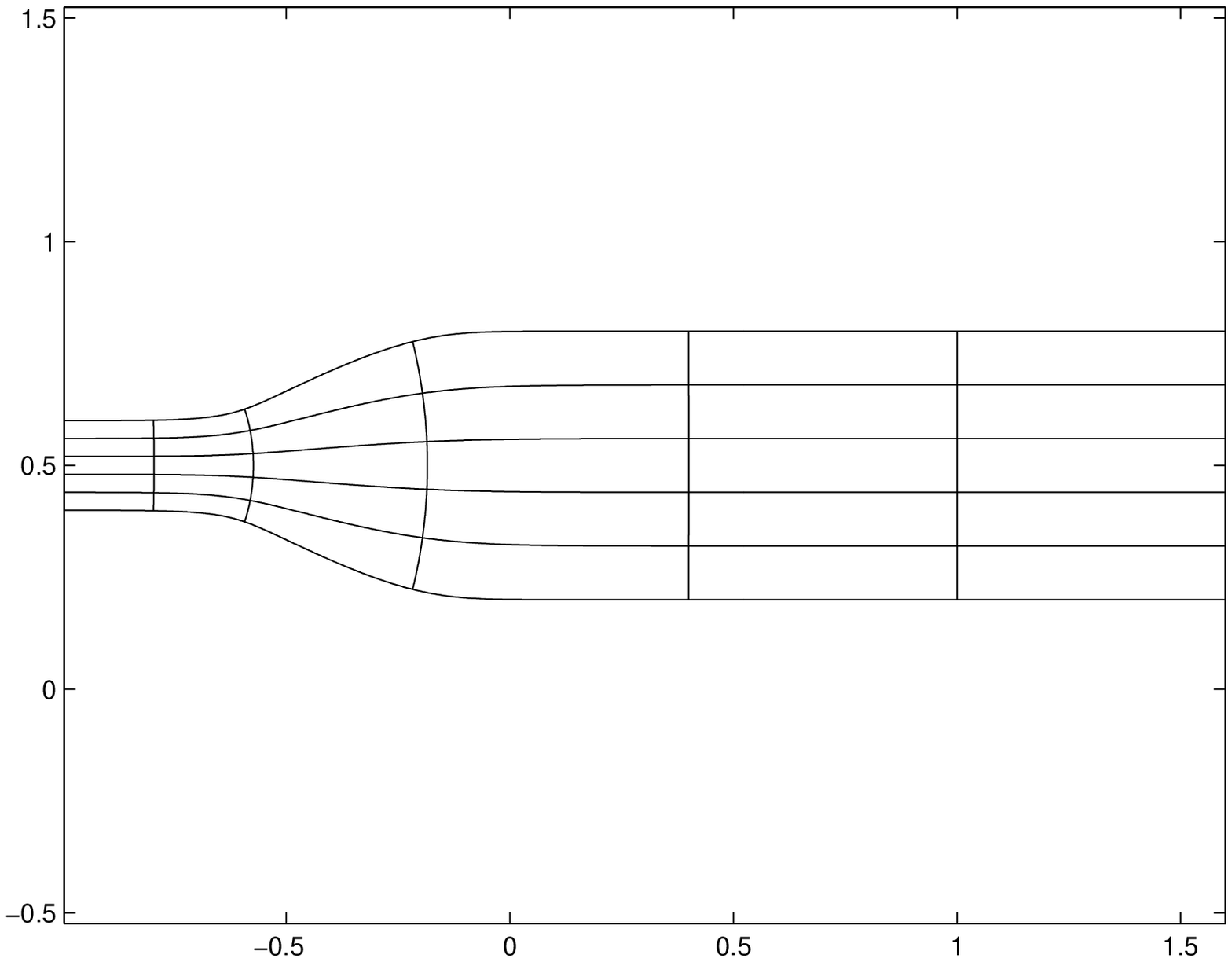}
  \includegraphics[width=0.45\textwidth]{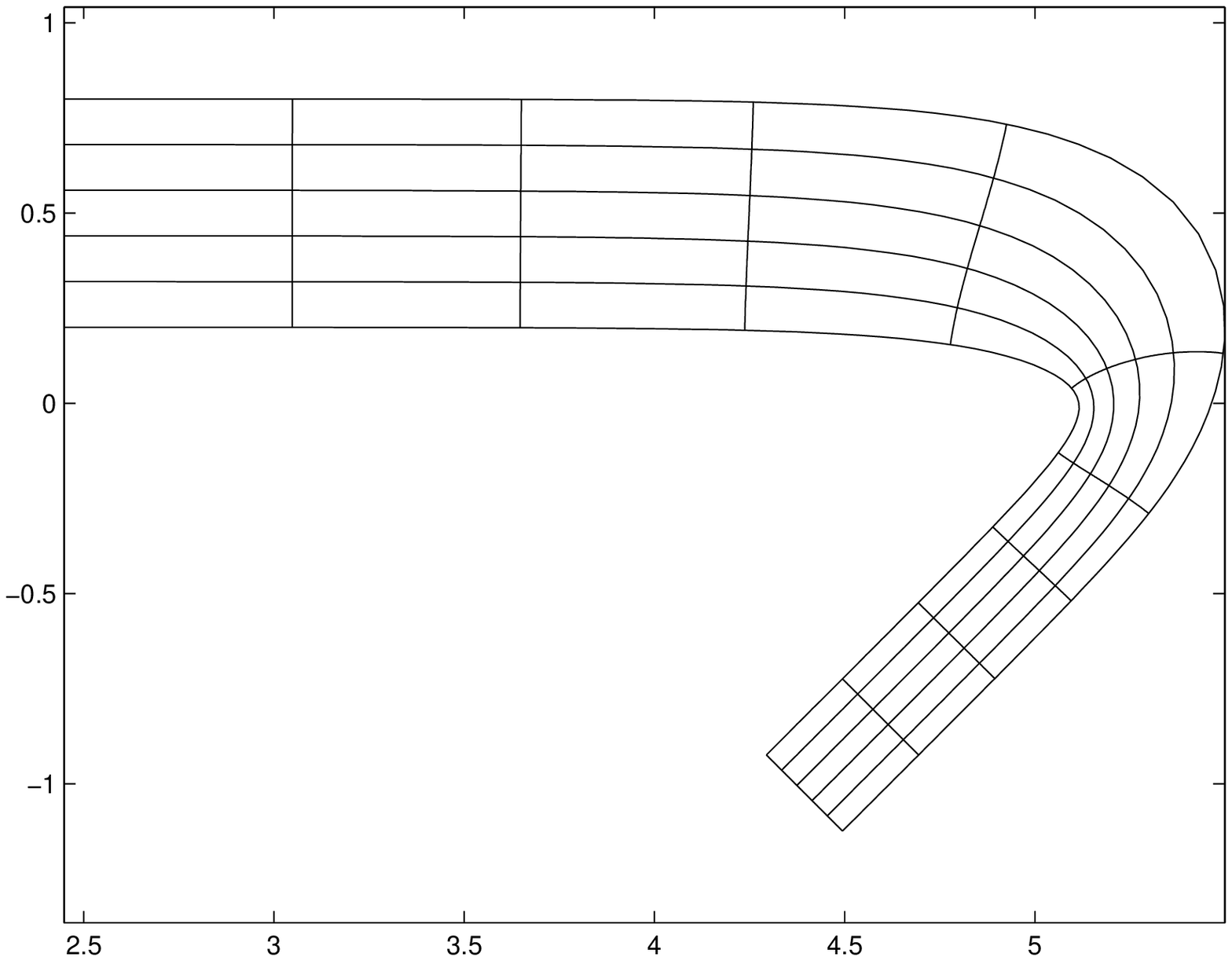}
  \caption{The two building blocks. The grid lines are images under the
    conformal mappings of $u=-5,-4,...,4,5$ and $v=0,0.2,\dots,1$.}
  \label{fig:sc}
\end{figure}

For $\Omega_1$, shown in Fig.~\ref{fig:sc}(a), a conformal mapping is
constructed using the approximate curve factor technique developed in
\cite{andersson-acf:2009}. The conformal mapping is $F_1=f_1\circ
g_1$, where
\begin{equation}
  \label{eq:scconfmap1}
  f_1(w)=A\int_{w_0}^w
  \prod_{j=1}^4\left(
    \sqrt{(\omega+b_k\i-w_k)^2-c_k^2}-b_k\i
  \right)^{\alpha_k-1}\omega^{-1}d\omega+z_0,
\end{equation}
and
\begin{equation}
  \label{eq:scconfmap2}
  g_1(w)=\exp(\pi w).
\end{equation}
In (\ref{eq:scconfmap1}), $A=0.6/\pi$ to get the width $0.6$ to the
right, $\alpha_1=\alpha_4=0.85$ and $\alpha_2=\alpha_3=1.15$ to get
inner angles of sizes $1.15\pi$ and $0.85\pi$, $b_1=b_4=c_1=c_4=1$ and
$b_2=b_3=c_2=c_3=0.05$ to get the corners appropriately rounded, while
$w_1=-1$, $w_2=-a$, $w_3=a$ and $w_4=1$, where $a=0.008740$ has been
numerically
determined to get the width $0.2$ to the left. Finally, $w_0$ is set
to $2$ and $z_0$ to $1+0.2i$ to position the waveguide in the complex
plane.

For $\Omega_3$, shown in Fig.~\ref{fig:sc}(b), the outer polygon
method \cite{andersson-outpol:2008} is used to construct the conformal
mapping. The mapping function is here $F_2=f_2\circ g_2$, where
\begin{equation}
  \label{eq:bcconfmap1}
  f_2(w)=A\int_{g_2(w_0)}^w\frac{(\omega-1)^{\alpha-1}}
  {(\omega+1)^{\alpha-1}(\omega-a)}\,d\omega+z_0
\end{equation}
and
\begin{equation}
  \label{eq:bcconfmap2}
  g_2(w)=w^{(\phi_2-\phi_1)/\pi}\e^{\i\phi_1}+a,
\end{equation}
with $A=0.1501\exp(3\pi\i/4)$, $\alpha=7/4$, $\phi_1=3\pi/10$,
$\phi_2=7\pi/10$, $a=-0.4632$, $w_0=-7$ and $z_0=4.4485+0.2\i$.

As has been mentioned in Section~\ref{sec:oneblock}, the blocks are in
theory assumed to be of infinite length. When determining the
reflection and transmisson matrices for one block, it must have
sufficiently long straight parts in both ends. The requirement is here
that the derivative of the conformal mapping must be practically
constant. In this example, it is sufficient to use $-5\le u\le5$ in
$\Omega_1$ and $-7\le u\le7$ in $\Omega_3$.  Hence, in the
calculations, $\Omega_1=F_1([-5,5]+i[0,1])$ and
$\Omega_3=F_2([-7,7]+i[0,1])$.

\subsection{Determination of the field and of reflection and
transmission matrices}
\label{sec:determ-field-refl}

The fields inside $\Omega_1$ and $\Omega_3$ have been determined using
the techniques described in Section~\ref{sec:DtN}. Simultaneously,
reflection and transmission operators for $\Omega_1$ and $\Omega_3$
have been determined using the techniques in Section~\ref{sec:RT}. All
calculations have been made using truncated matrices in the
differential equations (\ref{eq:Ricatti1})--(\ref{eq:Ricatti4}) and
(\ref{eq:DtNdef1}--\ref{eq:DtNDE2}) and a standard numerical ODE
solver (\verb+ode45+).

We have assumed a source at infinity to the left resulting in a
right-marching wave $\vec\Phi_{\text{in}}=(1\ 0\ 0\ 0\dots)^t$
entering the waveguide from the left. No sources to the right are
assumed.

The matrices $A(u)$ and $B^2(u)$ in (\ref{eq:DE}), as well as the
matrices $J$, $K$, $L$ and $M$ in (\ref{eq:alphabeta}) have been
determined for $u=-5,-4.99,...,5$ in $\Omega_1$, and for
$u=-7,-6.99,...,7$ in $\Omega_3$. Linear interpolation was then used
in the ODE solvers to determine $J$, $K$, $L$ and $M$ in
eqs. ~(\ref{eq:Ricatti1}--\ref{eq:Ricatti4}) and $A$ and $B^2$ in
eqs.~(\ref{eq:DtNDE1}) and (\ref{eq:DtNDE2}) for $u$ values not in
this set.

Finally, the field in $\Omega_2$ as well as reflection and
transmission matrices for the whole waveguide where calculated using
the Building Block method described in
Section~\ref{sec:comb-blocks-build}.

\subsection{Results}
\label{sec:results}

\begin{figure}[t]
  \centering
  \includegraphics[width=\linewidth]{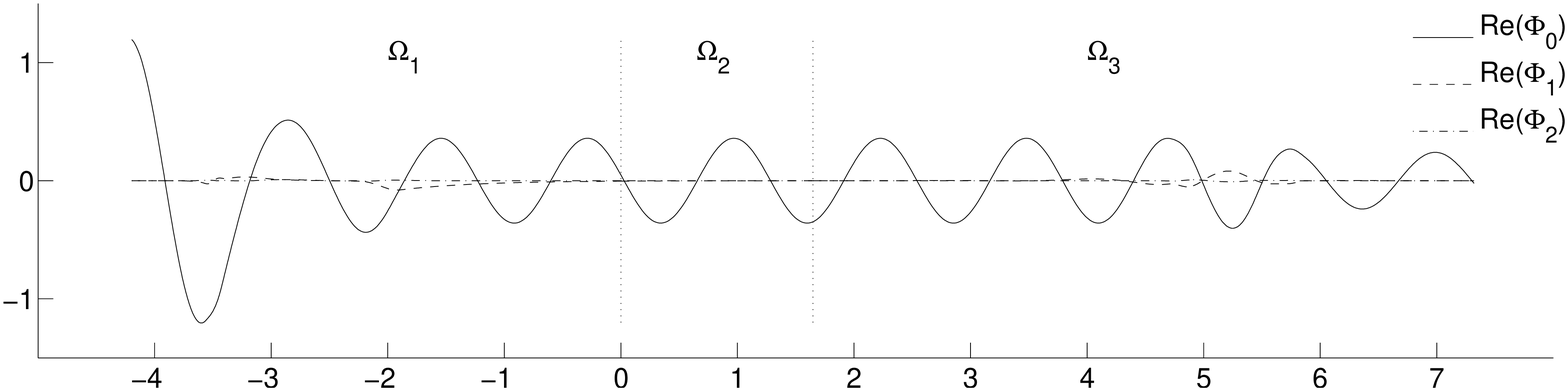}
  \includegraphics[width=\linewidth]{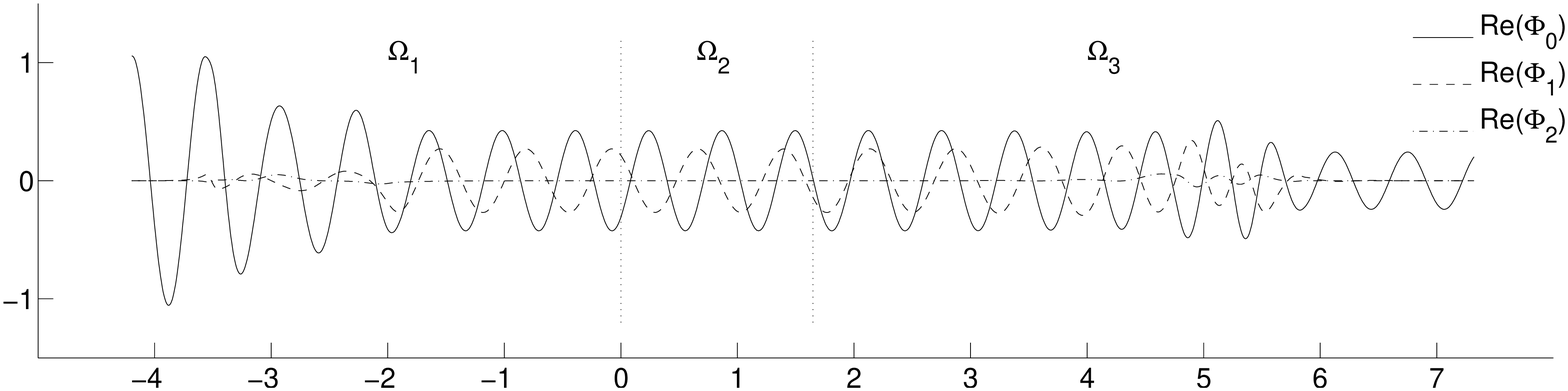}
  \includegraphics[width=\linewidth]{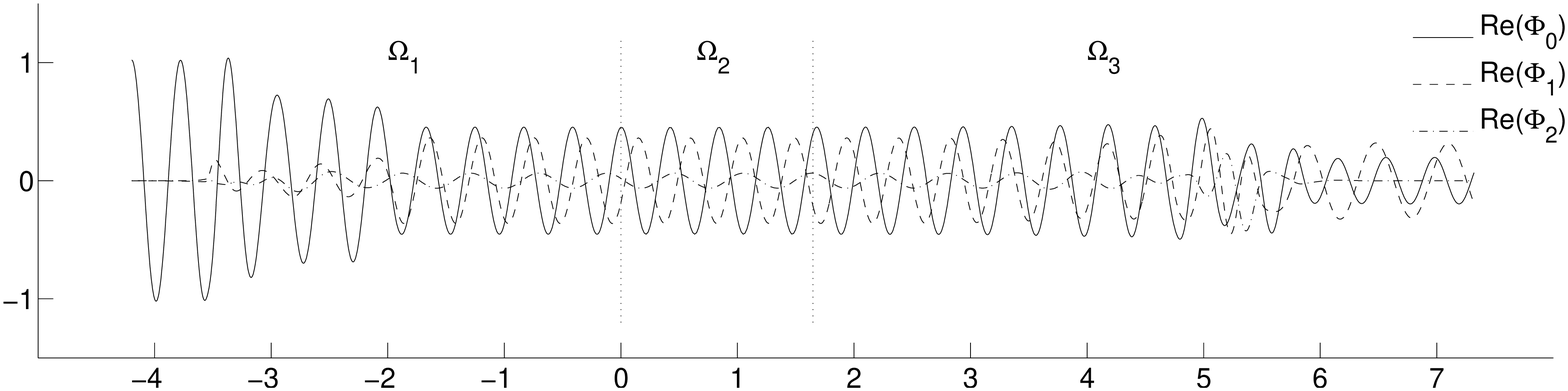}
  \caption{$\Re\Phi_0$, $\Re\Phi_1$ and $\Re\Phi_2$, are plotted for
    $k=5$, $k=10$ and $k=15$ for the incident wave $\Phiin=(1\ 0\ 0\
    0\ 0\ 0\ 0\ 0\ 0\ 0)^t$. The measure on the horizontal axis is the
    distance along the central curve in the waveguide, coinciding with
    the $x$--axis in Figure \ref{fig:exwg} for the interval
    (-4,7). Dotted vertical lines indicate borders between $\Omega_1$,
    $\Omega_2$ and $\Omega_3$. For $k=5$ ($ka=1$), only $\Phi_0$ is
    propagating (without attenuation) in the three straight parts of
    the waveguide, while for $k=10$ ($ka=2$), $\Phi_1$ propagates in
    $\Omega_2$ but not in the straight part of $\Omega_3$. For $k=15$
    ($ka=3$) finally, all three modes are propagating in $\Omega_2$
    and $\Phi_0$ and $\Phi_1$ propagate in the straight part of
    $\Omega_3$.}
  \label{fig:phi}
\end{figure}

The methods have been applied for $0<ka\le4$ using $10\times10$
matrices, where $a=0.2$ is the width of the waveguide at the left
end. The real part of the three lowest modes inside the waveguide in
Figure~\ref{fig:exwg} is shown in Figure~\ref{fig:phi} for the
incident wave $\Phiin=(1\ 0\ 0\ 0\ 0\ 0\ 0\ 0\ 0\ 0)^t$. For the
lowest frequency, only the lowest mode is present except near the
waveguide modifications, where also the second mode is visible. For
higher frequencies, the situation is more complicated with more
propagating modes.

\begin{figure}[t]
  \centering
  \includegraphics[width=0.9\linewidth]{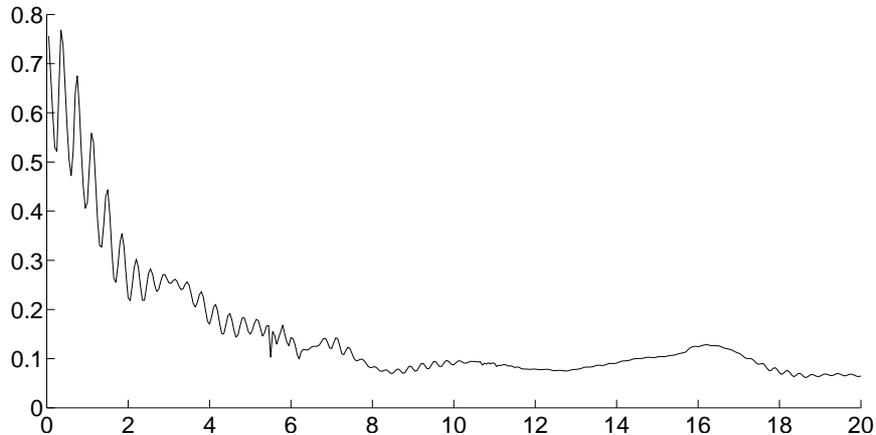}
  \caption{Power ratio $\mathcal P$ as function of $k$}
  \label{fig:ekvot}
\end{figure}

Of special interest is the power lost at the boundary where the
admittance has a non-vanishing real part. The power of mode $\Phi_n$
with complex amplitude $A_n$ in a waveguide with width $a$ is
proportional to
\begin{equation}\label{eq:power}
  \begin{cases}
    \frac{a(1+\delta_{n0})}{2k}\abs{A_n}^2
    \sqrt{k^2-\left(\frac{n\pi}a\right)^2},\quad&
    k\ge\frac{n\pi}a,\\
    0,&\text{otherwise}.
  \end{cases}
\end{equation}
The power ratio $\mathcal P$ for the waveguide is defined as the sum
of transmitted and reflected power over incident power. $\mathcal P$
is less or equal to 1 for a passive structure and less than 1 for
passive structure wit´h attenuation. In Figure~\ref{fig:ekvot},
$\mathcal P$ is shown as a function of $k$. For the chosen frequency
independent normal admittance there are losses present for all
frequencies, at least 25 \% in the low frequency end stabilizing at
about 90 \% for $k$ larger than 8. This corresponds to a total
attenuation of about 1 dB and 10 dB, respectively.

\subsubsection{Comparison between the RT and DtN methods}
\label{sec:RT-DtN}

In this section, the RT and DtN methods are compared. The DtN method
described in Section~\ref{sec:DtN} is used in $\Omega_1$ and
$\Omega_3$, whereas the reflection and transmission matrices from the
RT method are used in the BB method, described in Section 3.2.1, for
$\Omega_2$ and is an alternative to the DtN method for the right end
of $\Omega_3$. Both methods should give identical results at the
border between $\Omega_1$ and $\Omega_2$ and at the right end of
$\Omega_3$. For the first case, $\vec\Phi$ using the DtN method is
compared with $(C^++C^{-−})\vec\Phiin$. Likewise, for the second case
$\vec\Phi$ with the DtN method is compared with $\Ttot\Phiin$.  Here,
$C^\pm$ is determined using (47) and $\Ttot$ using (48).

The smooth curves visualized in Figure~\ref{fig:phi} indicates that
for the three $k$ values presented there, the correspondence is
good. Similar results are obtained all over the investigated interval
$0<k\le20$, no matter if only one or if several propagating modes are
present at the two points in the waveguide. One example of a more
detailed comparison is given in Table~\ref{tab:phires-bcend}, where
results for $k=15$ are compared.

\begin{table}[ht]
  \begin{equation*}
    \begin{array}{|l|rrl|}
      \hline
      &\multicolumn1c{\vec\Phi_{\Omega_1}(\text{end})}
      &\multicolumn1c{\vec\Phi_{\Omega_2}(0)}
      &\abs{\text{difference}}\\
      \hline
      \Phi_1&0.4521-0.0448\i&0.4521-0.0449\i&3.744\cdot10^{-5}\\
      \Phi_2&-0.1873-0.2693\i&-0.1873-0.2693\i&7.971\cdot10^{-6}\\
      \Phi_3&0.0190+0.0203\i&0.0190+0.0203\i&7.730\cdot10^{-6}\\
      \hline
      &\multicolumn1c{\vec\Phi_{\Omega_3}(\text{end})}
      &\multicolumn1c{\Ttot\Phiin}
      &\\
      \hline
      \Phi_1&0.0671-0.1847\i&0.0671-0.1848\i&2.792\cdot10^{-5}\\
      \Phi_2&-0.1926+0.2528\i&-0.1926+0.2528\i&2.812\cdot10^{-5}\\
      \Phi_3&\multicolumn1c{\approx0}&\multicolumn1c{\approx0}&\\
      \hline
    \end{array}
  \end{equation*}
  \caption{Comparison of the RT and DtN method. Above: $\Phi_1$,
    $\Phi_2$ and $\Phi_3$ at the border between $\Omega_1$ and $\Omega_2$
    calculated with the two different methods. Below: $\Phi_1$ and
    $\Phi_2$ at the end of $\Omega_3$ calculated with the two
    different methods. All calculations are made for $k=15$ ($ka=3$)
    and with $10\times10$ matrices.}
  \label{tab:phires-bcend}
\end{table}

\subsubsection{Comparison with a FE solution}
\label{sec:FEMvsFourier}

\begin{figure}[t]
  \centering
  \includegraphics[width=0.9\linewidth]{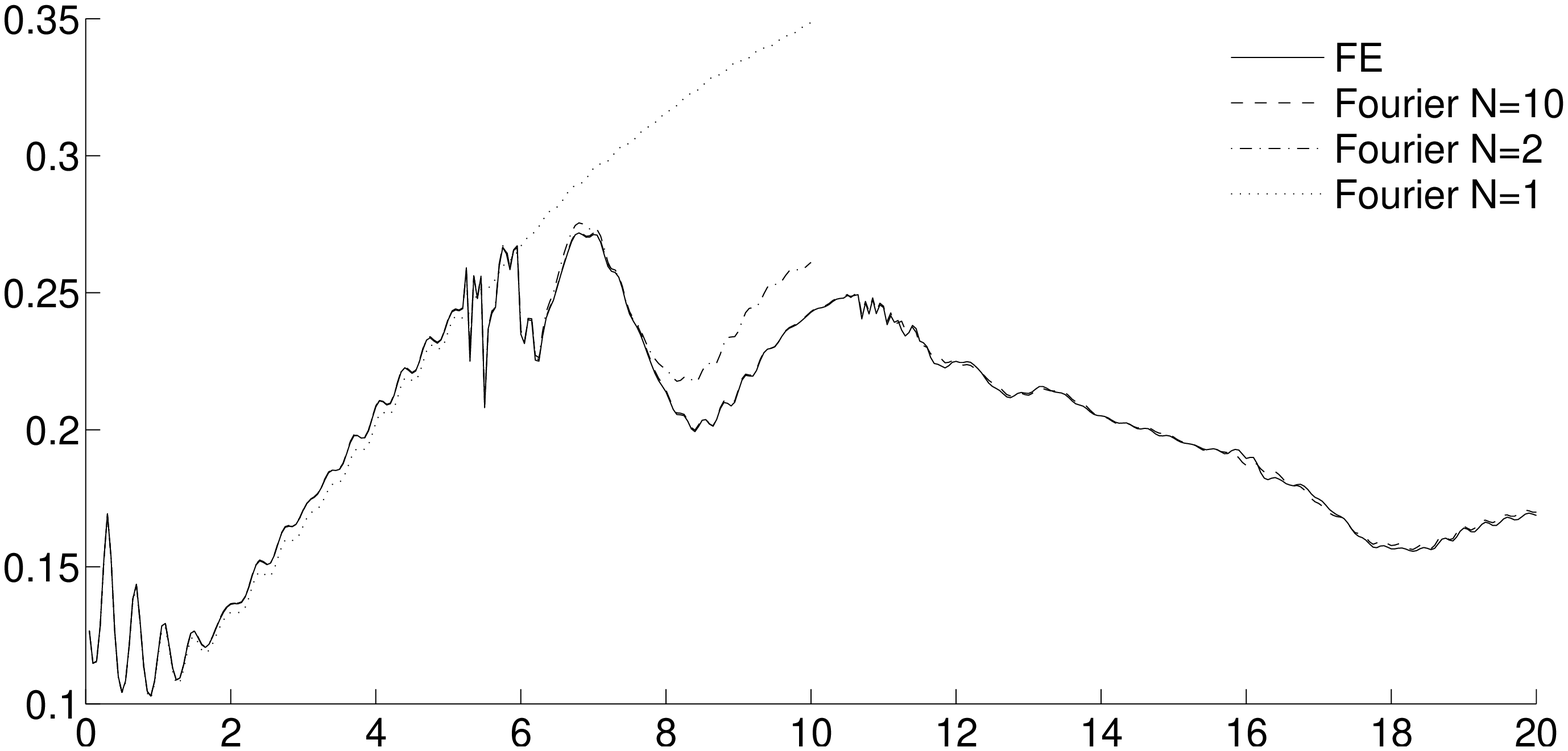}
  \caption{$\abs{\Phi_{0}}$ at the end of $\Omega_3$ calculated for
    $k=0.05,0.1,\dots,20$ with the Finite Element Method and with the
    Fourier methods described in the article with $N=1,2$ and $10$,
    i.e., using $1\times1$, $2\times2$ and $10\times10$ matrices in
    the calculations.}
  \label{fig:FEMvsFourier}
\end{figure}

\begin{figure}[b]
  \centering
  \includegraphics[width=0.9\linewidth]{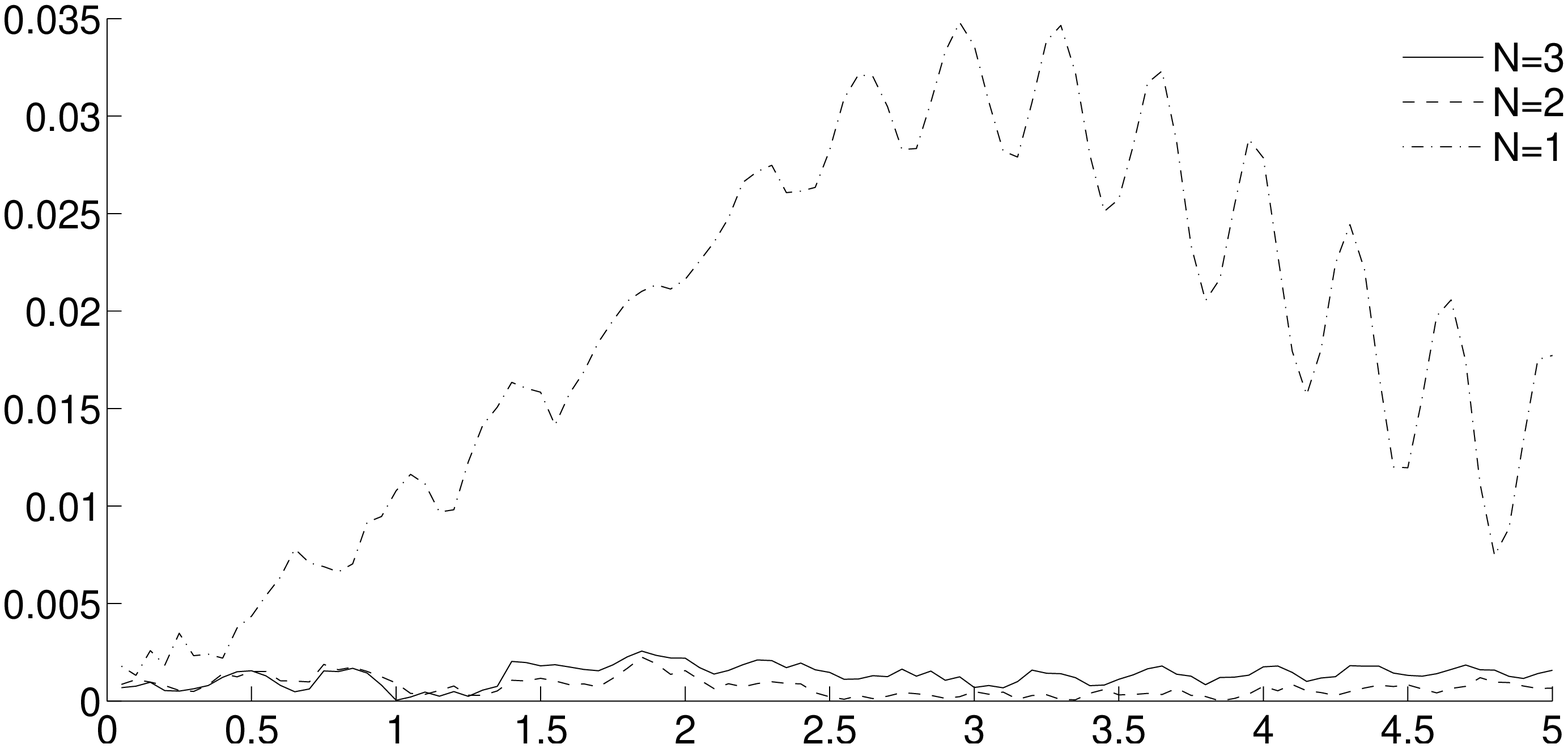}
  \caption{$\abs{\abs{\Phi_{\text{Four}}/\Phi_{\text{FEM}}}-1}$, where
    $\Phi_{\text{FEM}}$ and $\Phi_{\text{Four}}$ are $\Phi_0$ at the
    end of $\Omega_3$ calculated for $0<k\le5$ with FE software and
    Fourier methods respectively. The calculations of
    $\Phi_{\text{Four}}$ are done with $N=3,2$ and $1$, i.e., with
    $3\times3$, $2\times2$ and $1\times1$ matrices.}
  \label{fig:relerr123}
\end{figure}

\begin{figure}[t]
  \centering
  \includegraphics[width=0.9\linewidth]{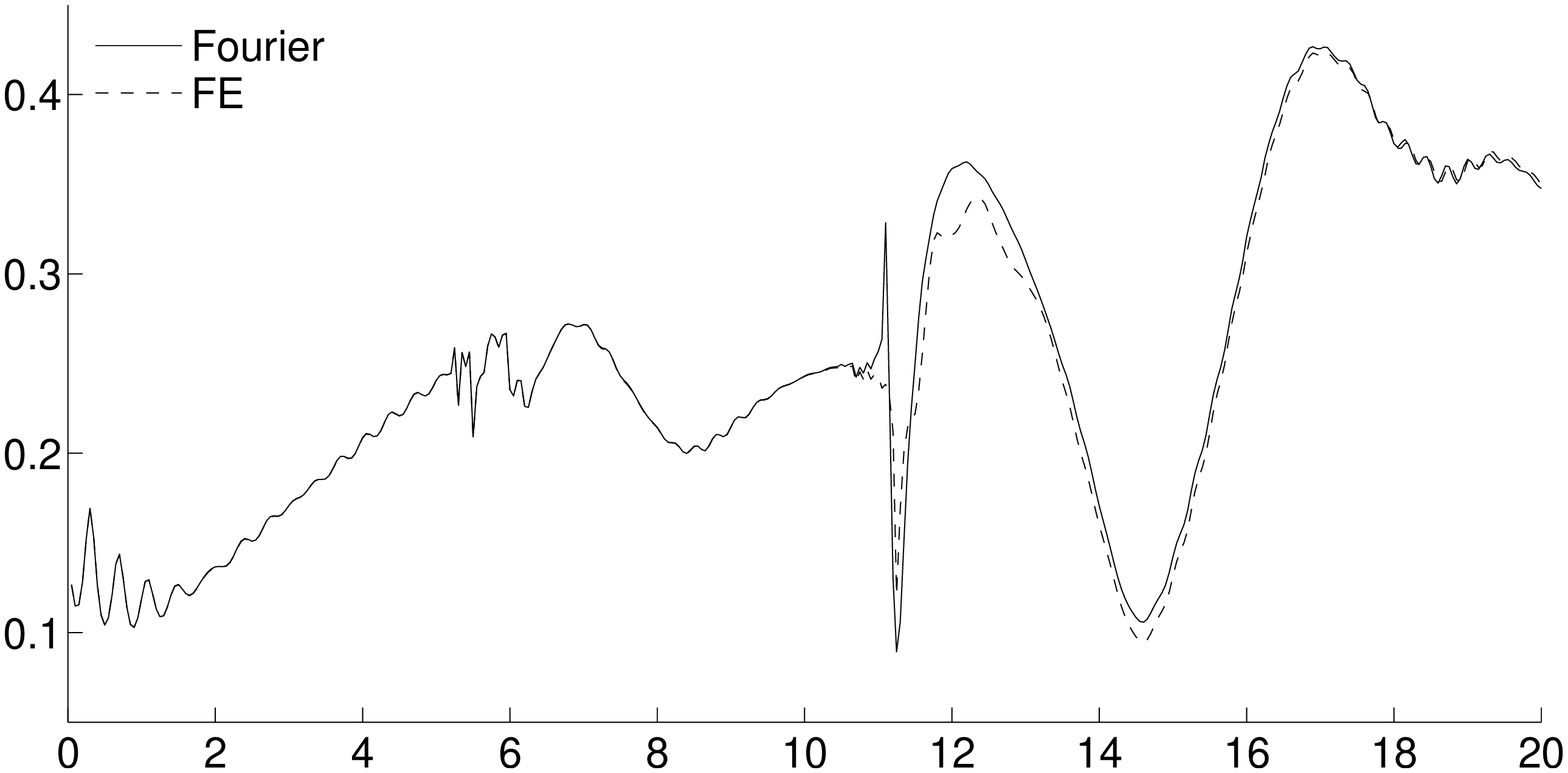}
  \caption{The total field $\abs\phi$ in the point $(7.0671,-1.5855)$
    (the left point at the end of the waveguide) for $0<k<20$
    calculated with Fourier methods using $N=10$ and with FE
    software.}
  \label{fig:FEMvsFour2}
\end{figure}

\begin{figure}[b]
  \centering
  \includegraphics[width=0.9\linewidth]{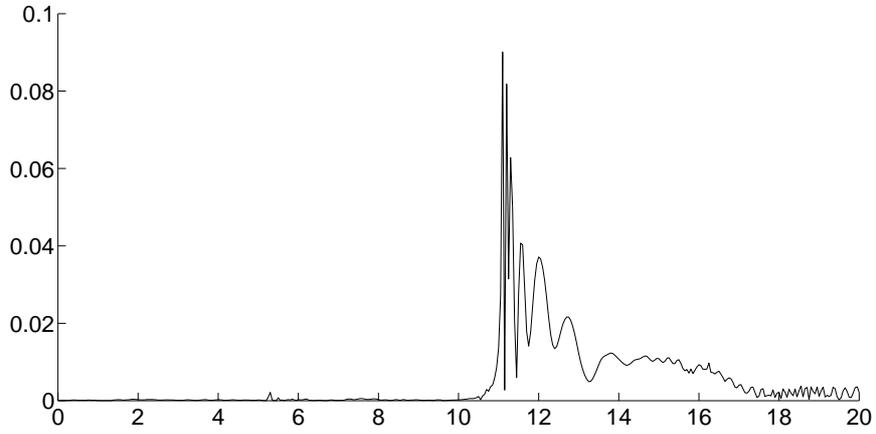}
  \caption{The absolute difference
    $\abs{\abs{\varphi_{\text{Four}}}-\abs{\varphi_{\text{FE}}}}$ for
    $0<k<20$ in the point $(7.0671,-1.5855)$, where
    $\varphi_{\text{Four}}$ and $\varphi_{\text{FE}}$ are $\varphi$ calculated
    with Fourier methods and FE software respectively.}
  \label{fig:totalfaltFEMFour}
\end{figure}

As a comparison, we have applied a commercial FE solver (COMSOL
Multiphysics) to our model problem. In
Figures~\ref{fig:FEMvsFourier}--\ref{fig:totalfaltFEMFour}, FE
solutions for $0<k\le20$ are compared with the corresponding Fourier
method solutions. From the FE solution, it is easy to extract the base
mode $\Phi_0$ by calculating the average field over a cross-section of
the waveguide. In Figure~\ref{fig:FEMvsFourier}, the FE calculated
$\Phi_0$ on the boundary at the end of the waveguide is compared with
$\Phi_0$ calculated with Fourier methods using three different matrix
sizes.

For low frequencies, Figure~\ref{fig:FEMvsFourier} indicates that
accurate results can be achieved with small matrices. The situation is
examined further in Figure~\ref{fig:relerr123} where a relative
difference between the FE solution and Fourier solutions using
$1\times1$, $2\times2$ and $3\times3$ matrices are shown.

The width of the waveguide is $0.6$ in the central part, which means
that there, the second mode $\Phi_1$ is propagating (without
attenuation) for $k>\pi/0.6\approx5.24$. Clearly, calculations with
$1\times1$ matrices can not give accurate results for $k>\pi/0.6$.
However, it is interesting to note that even when using only one
matrix element in the calculations, fairly good results are achieved
for frequencies with not more than one propagating mode. An important
feature of the Fourier method is its ability for low frequencies to
partition the wave field into a low number of left and right going
waves adding valuable physical understanding.

For $k>\pi/(0.2\sqrt2)\approx11.1$, i.e. when the second mode is
propagating without attenuation in the end of the waveguide, the
differences are significantly greater, see
Figures~\ref{fig:FEMvsFour2} and \ref{fig:totalfaltFEMFour}. Note
however that for $k$ just above the frequency $11.1$, there are
significant interference effects between the two propagating
modes. For these frequencies, the problem is ill-posed since the
interference is very sensitive to small changes in not only $k$ but
also in geometry and boundary conditions. Due to that there are
inevitable small differences in the two models, the discrepancies
shown in Figures~\ref{fig:FEMvsFour2} and \ref{fig:totalfaltFEMFour}
are therefore not surprising. The main reason for this ill-posedness
is that the axial wavenumber of the second mode is small increasing
from zero when it starts propagating. Note also that the differences
tend to vanish for higher frequencies.  In $\Omega_3$, there are only
forward-marching waves present. Additionally, due to the Neumann
boundary condition at the end of $\Omega_3$, the first mode is
constant over the cross-section, while the second mode is at maximum
at the boundary. Therefore, the interference is large at the chosen
point. At points inside the waveguide, the differences between the two
methods are much smaller.

\begin{figure}[t]
  \centering
  \includegraphics[scale=0.45]{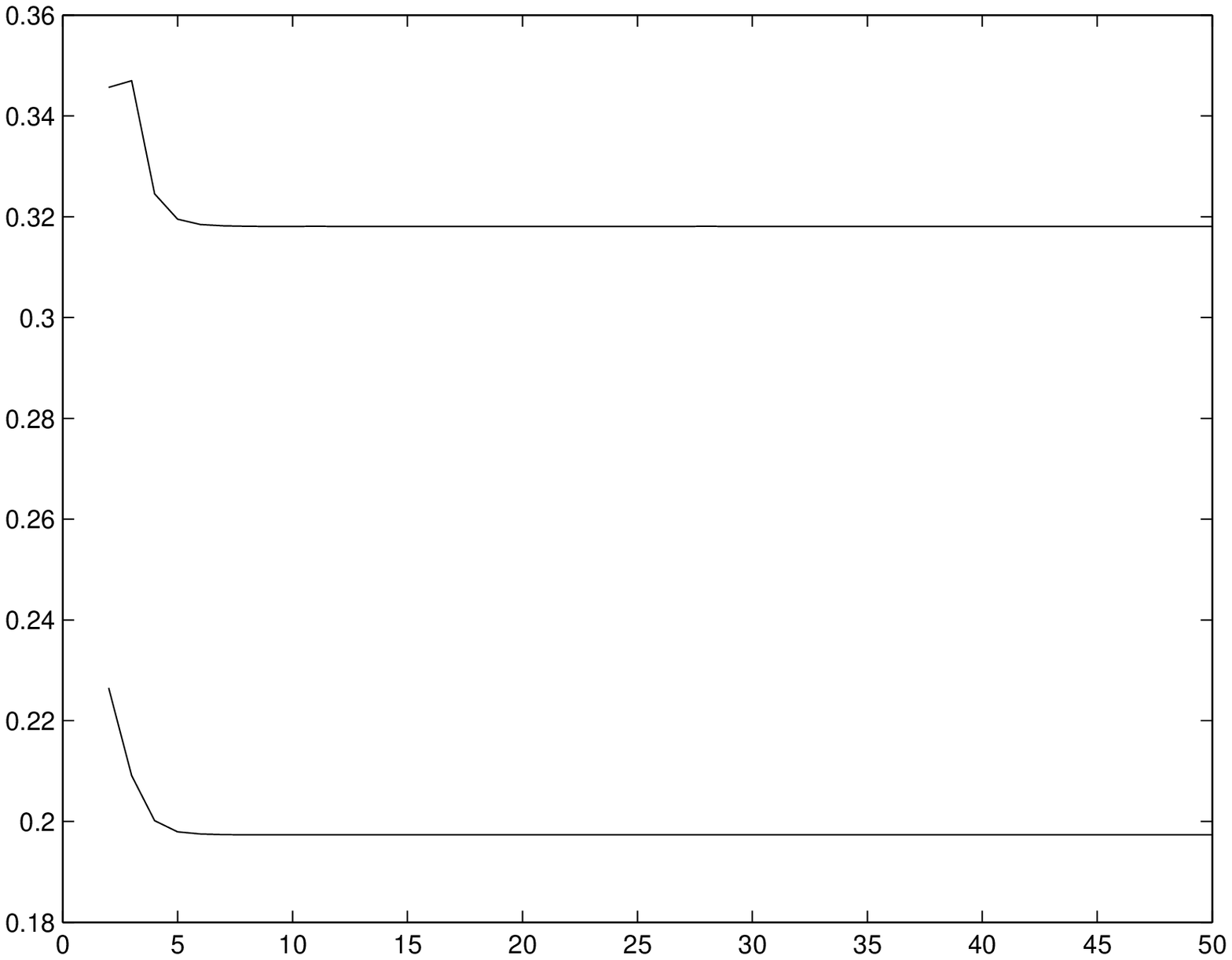}
  \caption{$\Phi_0$ and $\Phi_1$ at the end of the waveguide
    calculated for $k=15$ using matrices of size $N\times N$ with
    $N=2\dots50$.}
  \label{fig:T_N}
\end{figure}

For higher frequencies in the investigated interval, an indication of
the required matrix size is given in Figure~\ref{fig:T_N}, where the
problem is solved for $k=15$ using matrix sizes from $2\times2$ upto
$50\times50$. The results are stable to three significant figures for
matrix sizes $7\times7$ and upwards.

\section{Discussion and conclusion}
\label{sec:conclusion}

The example problem in Section~\ref{sec:numerical-example} is of
course not a ``general'' wave\-guide. There are numerous possible
boundary variations not commented so far. For example, the methods in
this article seem to require smooth changes in both geometry and
boundary conditions to get converging Fourier series. However, the
Building Block method and well-established mode-matching techniques
can overcome most such problems. Furthermore, as is illustrated in
\cite{Nilsson:2002} where an L-bend is investigated, good results can
be achieved even when the conformal mapping functions contain
singularities on the boundary. It is however evident that the
differential equations get stiffer and that larger truncated matrices
are required.

Another simplification in this article is the assumption of a Neumann
boundary condition on one of the boundaries. An iterated use of the
Building Block method and lengthways partitions of the waveguide can
handle two-dimensional waveguides with non-hard walls on both sides.

As a reference and comparison, the problem in the example has been
solved using commercial software for the finite element method.
As was seen in
Figures~\ref{fig:FEMvsFourier}--\ref{fig:totalfaltFEMFour}, the
correspondence between a FE solution and the Fourier methods solution
is good, with a small tendency to growing discrepancy with growing
$k$. This is not surprising; to maintain a certain accuracy when $k$
increases, both methods require enhanced numerical work. In FE, a
finer mesh is needed, for details see for example
\cite{Ihlenburg:1998}, while the Fourier methods require larger
matrices.

When working with $10\times10$ matrices, the Fourier method described
in this article is very slow compared to the FE method. To produce the
results presenterd in Figure \ref{fig:FEMvsFourier}, i.e., to solve
the problem for 400 different $k$ values, takes between one and two hours
with the FE method on an ordinary PC, while the copmputation
time for the Fourier method was several days on the same machine. In
the FE method, the mesh was set fine enough to give accurate results
in the higher parts of the investigated frequency interval.

It should of course be mentioned that no professional computer
programmer has been involved in writing the code for the Fourier
routines. We expect that there are still many possibilities to optimize
these programs for increased speed.

Note also that
for the greater part of the investigated frequency interval $0<k\le20$
in our model example, accurate results are achieved with matrix sizes
smaller than $10\times10$. Indeed,
truncating the matrices so that they contain only one
element gives fairly good results in intervals where only one
mode is propagating. And when using $1\times1$ matrices, even
unprofessional computer programs perform well.

The most time-consuming part of the calculations is the determination
of the matrices $A$ and $B^2$ in (\ref{eq:DE}) for a large set of
$u$-values. For every value of $u$, $\lambda_n(u)$ for $n=0,\dots,N-1$
should be found by solving equation (\ref{eq:lambdaeq}) numerically.
The values of $\alpha$ and $\beta$ in eqs.~(\ref{eq:vsinlambda}) and
(\ref{eq:vcos^2lambda}) can be determined analytically, but for every
$u$, $N^2$ numerical integrations are needed to determine the values
of $\mu_{mn}(u)$ in Eq.~(\ref{eq:mucoslambda}). For values of $u$,
corresponding to hard boundaries, $\lambda_n=n\pi$ and the $\mu$
coefficients can be calculated using Fast Fourier Transforms, but for
values of $u$ corresponding to boundaries with admittance, a
comparatively slow numerical integration must be used for each matrix
element.

A possible drawback for the FE method is that it needs more computer
memory in order to determine the wave field for higher frequencies.
If we consider an object that consists of a couple of complicated
blocks that are in arbitrary order repeatedly connected with straight
waveguides, the Fourier method could be an attractive alternative when
considering computation time.

Another drawback for some FE commercial packages, not present in the
Fourier method, is the treatment of low frequency
electromagnetic waves. This is a (solvable)
\cite{AndriulliCoolsOlyslagerMichielssen2008,chew2014} deficiency.

For time harmonic wave scattering problems, the Fourier methods
described in this article are applicable for the low and medium
frequency domain. It is beyond the scope of this paper to develop more
precise bounds for this domain. The method adds physical
interpretation, in particular at low frequencies, like partition of
the wave field into left and right going physical modes and
facilitates a power loss analysis also into terms of modes.

In summary, we conclude, based primarily on requirements from
industry, that more than one type of time harmonic waveguide
simulation tool is required. It is demonstrated that Fourier
methods based on Fourier Analysis provides one such tool. Its accuracy
is checked against FE simulations for a general two-dimensional
waveguide with normal admittance boundary conditions at low and medium
frequencies. For the current investigation with non-zero normal
boundary admittance, the Fourier method, with its present
implementation, is considerably slower than the FE method that is more
memory demanding. However, for inverse engineering involving tuning of
straight waveguides, the Fourier method is an attractive
alternative including time aspects. The prime motivation for the
Fourier method is its added physical understanding primarily
at low frequencies.

\bibliographystyle{plain} 
\bibliography{referenser}

\end{document}